\documentclass[onecolumn,amsmath,amssymb,nofootinbib,12pt]{article}
\pdfoutput=1 

\usepackage{jheppub} 

\usepackage[multiple]{footmisc}

\usepackage{tensor}
\usepackage{graphicx,subcaption}
\graphicspath{ {figures/} }	
\usepackage{amsfonts}
\usepackage{xparse}
 \usepackage{amsmath}
 \usepackage{amssymb}
 \usepackage{mathtools}
\usepackage{epsfig}
\usepackage{pdfpages}
\usepackage{bbm}
\usepackage{graphicx,epstopdf}
\usepackage[numbers]{natbib}
\usepackage[makeroom]{cancel}
\usepackage{hyperref}
\usepackage{array}
\usepackage[export]{adjustbox}
\usepackage[normalem]{ulem}
\usepackage{tcolorbox}
\usepackage{romannum}

\title{Revisiting thermodynamic topologies of black holes}

\author[a]{Chaoxi Fang,}
\author[b]{Jie Jiang,}
\author[a]{Ming Zhang \footnote{Corresponding author}}

\affiliation[a]{Department of Physics, Jiangxi Normal University,\\ Nanchang 330022, China}
\affiliation[b]{College of Education for the Future, Beijing Normal University, \\ Zhuhai 519087, China}

\emailAdd{chaoxi.f@jxnu.edu.cn}
\emailAdd{jiejiang@mail.bnu.edu.cn}
\emailAdd{mingzhang@jxnu.edu.cn}

\date{\today}

\abstract{In the generalized off-shell free energy landscape, black holes can be treated as thermodynamic topological defects. The local topological properties of the spacetime can be reflected by the winding numbers at the defects, while the global topological nature can be classified  by the topological number which is the sum of  all local winding numbers. We propose that the winding numbers can be calculated via the residues of isolated one-order  pole points  of   characterized  functions constructed from the off-shell free energy. Using the residue method, we show that the topologies of black holes can be divided into three classes with the topological numbers being -1, 0, and 1, respectively, being consistent with the results obtained in [Phys. Rev. Lett. 129, 191101 (2022)] by using the topological current method. Moreover, we point out that standard defect points, generation and annihilation points, and critical points can be distinguished by coefficients of the Laurent series of the off-shell characterized function at those singular points.
}

\begin{document}

\maketitle

\section{Introduction}
The topological defect is a universal pattern of nature and plays a fundamental role in providing stability information against perturbation for systems from liquid crystals to particle physics \cite{Fumeron:2022gcd}. The study of topological properties in strong gravity systems beyond in quantum many-body systems can be traced back to \cite{Cunha:2017qtt,Cunha:2020azh,Guo:2020qwk}, where the existence of at least one light ring around the four-dimensional stationary non-extremal rotating black hole was proved. Recently, exploring thermodynamics of black holes via topological properties has charted a way for deepening our recognition of intrinsic characteristics of spacetime \cite{Wei:2021vdx,Yerra:2022alz,Fan:2022bsq,Yerra:2022coh,Yerra:2022eov}. The topologies merit neglecting the details of objects and just concerns with their generic properties. To investigate the thermodynamic topology properties of black holes as solutions of Einstein gravitational field equation, one has to identify the solutions as topological thermodynamic defects in the thermodynamic parameter space which can be constructed in the generalized off-shell free energy landscape \cite{Wei:2022dzw}.

In the standard definition of free energy, the Hawking temperature is invariant for a canonical ensemble containing states of black holes with various radii. The generalized concept of free energy was raised in Ref. \cite{York:1986it} and was recently used to study the thermodynamics of the five-dimensional Schwarzschild black hole in Ref. \cite{Andre:2020czm}. In the generalized landscape of free energy, a canonical ensemble containing different black hole states characterized by the horizon radii is endowed with an ensemble temperature fluctuating around the Hawking temperature. As a result, in the generalized free energy landscape, one has on-shell free energy states when the ensemble temperature equals to the Hawking temperature and off-shell free energy states when the ensemble temperature deviates from the Hawking temperature. The former one saturates the Einstein field equation whilst the latter one violates it \cite{Liu:2022aqt}. Notice, however, that the extremal points of the free energy can recover the on-shell states as the thermodynamic first law of the black hole is satisfied there. Locally, a minimum (maximum) point on the landscape of free energy corresponds to a stable (unstable) thermodynamic state of the black hole, as the specific heat there is positive (negative). Globally, thermodynamically stable black hole occupies a minimal point in the landscape of free energy \cite{Hawking:1982dh,Kubiznak:2012wp}. Researches along this line include Ref. \cite{Li:2022oup} for the proposal of a generalized free energy landscape for the Anti-de Sitter (AdS) black holes and Refs. \cite{Li:2020khm,Li:2020nsy,Liu:2021lmr} for the dynamical phase transition of spherically symmetric black holes.

The aim of this paper is to propose an alternative way to investigate the thermodynamic topologies of black holes beyond the topological current method. In \cite{Wei:2022dzw}, Duan's topological current method was introduced to calculate the winding number and the topological number of black holes as thermodynamic defects in the generalized free energy landscape. We observe that the winding numbers and topological numbers can be calculated by employing residues of the singular points of a characterized complex function constructed from the off-shell free energy. The method of residue can recover results obtained by the topological current method. Interestingly, we can also discern generation, annihilation, and critical points by coefficient analysis of the Laurent series of the off-shell characterized function. We arrange the paper as follows. In Sec. \ref{sec:gen}, we will give a brief review of the generalized landscape of free energy and introduce the residue method of investigating the local and global topological properties of the black hole solutions. In Sec. \ref{sec:ttbh}, we will apply the residue method to study the topological properties of Schwarzschild black holes, Reissner-Nordström (RN) black holes, RN-AdS black holes, and the black holes in non-linear electrodynamics. Sec. \ref{sec:con} will be devoted to our closing remarks.

\section{Generalized landscape of free energy}\label{sec:gen}
We shall take the partition function $\mathcal{Z}$ using the first-order Euclidean action $\mathcal{I}$ including a subtraction term \cite{Gibbons:1976ue},
\begin{equation}
\mathcal{Z}(\beta)=e^{-\beta F}=\int D[g] e^{-\frac{\mathcal{I}}{\hbar}} \sim e^{-\frac{\mathcal{I}}{\hbar}},
\end{equation}
where the Euclidean gravitational action $\mathcal{I}$ reads \cite{Li:2022oup}
\begin{equation}
\mathcal{I}=\mathcal{I}_1-\mathcal{I}_{\text {subtract }}=-\frac{1}{16 \pi} \int \sqrt{g} R d^4 x+\frac{1}{8 \pi} \oint \sqrt{\gamma} K d^3 x+\mathcal{I}_{\textrm{matt}}-\mathcal{I}_{\text {subtract }}.
\end{equation}
$\beta$ is the inverse of temperature or Euclidean time period, $K$ is the trace of extrinsic curvature tensor defined on the boundary with an induced metric $\gamma_{ij}$ and the determinant of this metric is $\gamma$. We use the saddle point approximation (semi-classical approximation) which allows us to evaluate the partition function on the Euclidean fluctuating black hole with fixed event horizon radius $r_+$ (which is the order parameter) and fixed ensemble temperature $T=\beta^{-1}$ (which is the temperature of the thermal environment and thus independent of the order parameter $r_+$ and adjustable). $\mathcal{I}_{\textrm{matt}}$ stands for the contribution from matter like the electromagnetic field. For the asymptotically AdS spacetime, the partition function can be evaluated just via the Einstein-Hilbert action
\begin{equation}
\mathcal{I}_E=-\frac{1}{16 \pi} \int\left(R+\frac{6}{L^2}\right) \sqrt{g} d^4 x+\mathcal{I}_{\textrm{matt}}
\end{equation} 
on the Euclidean gravitational instanton with the conical singularity \cite{Li:2022oup}, where $L$ is the AdS radius. The spacetime solution of the action turns asymptotically flat from AdS in the limit $L\to\infty$.

The thermodynamic energy and entropy of the system are respectively
\begin{equation}
E=\partial_\beta \mathcal{I},
\end{equation}
\begin{equation}
S=\left(\beta \partial_\beta \mathcal{I}-\mathcal{I}\right).
\end{equation}
Then the generalized free energy of the system can be defined as 
\begin{equation}
\mathcal{F}=E-\frac{S}{\tau},
\end{equation}
where the parameter $\tau$ owns the same dimension with the ensemble temperature $T$. Note that for the AdS black hole the thermodynamic energy $E$ is in fact the enthalpy $H$ \cite{Kubiznak:2016qmn,Kastor:2009wy,Chamblin:1999tk}. The on-shell free energy landscape is that 
\begin{equation}
\tau=\frac{1}{T},
\end{equation}
which renders that the black hole solution satisfies the Einstein equation
\begin{equation}\label{eineq}
\mathcal{E}_{\mu \nu} \equiv G_{\mu \nu}-\frac{8 \pi G}{c^4} T_{\mu \nu}=0,
\end{equation}
where $G_{\mu\nu}$ is the Einstein tensor of the spacetime background geometry, $T_{\mu\nu}$ is the energy-momentum tensor of matter. In the off-shell landscape where the field equation is violated, $\tau$ deviates from $T^{-1}$, which means that the temperature of the canonical ensemble fluctuates. 

As pointed out in \cite{Wei:2022dzw}, at the point $\tau=1/T$, we have 
\begin{equation}\label{dep}
\frac{\partial \mathcal{F}}{\partial r_{\mathrm{h}}}=0.
\end{equation}
Then a vector field $\phi$ is constructed there as \footnote{We also have $$\frac{\partial \mathcal{F}}{\partial S}=0,$$ so we can also construct a vector field as $$\phi=\left(\frac{\partial \mathcal{F}}{\partial S},-\cot \Theta \csc \Theta\right).$$}
\begin{equation}\label{phid}
\phi=\left(\frac{\partial \mathcal{F}}{\partial r_{\mathrm{h}}},-\cot \Theta \csc \Theta\right),
\end{equation}
where $0 \leq \Theta \leq \pi$ is an additional introduced parameter. Then using Duan's $\phi$-mapping topological current theory \cite{Duan:2018oup}, the local winding number of the topological defects at the points $\tau=1/T$ can be calculated, and the global topological number of the black hole solution is obtained by adding those winding number at each defect in \cite{Wei:2022dzw}. We can see that to do this calculation, an extension of the dimension of parameter space is needed and one should choose an auxiliary parameter $\Theta$ to construct a function $-\cot \Theta \csc \Theta$. Principally, the selection of the function can be arbitrary. The quintessential ingredient is to construct a plane so that a real contour can be introduced to enclose the thermodynamic defects defined by Eq. (\ref{dep}). 

This is reminiscent of the contour integral of a complex function $\mathcal{R}(z)$ which is analytic except at a finite number of isolated singular points $z_1, z_2, \ldots, z_m$ interior to $C$ which is a positively oriented simple closed contour. The complex function is defined within and on the contour $C$. According to the residue theorem,
\begin{equation}
\oint_C \frac{\mathcal{R}(z)}{2\pi i} d z= \sum_{k=1}^m \operatorname{Res}\left[\mathcal{R}\left(z_k\right)\right].
\end{equation}
Inspired by this, we first solve $\partial_{r_h}\mathcal{F}=0$
and obtain 
\begin{equation}
\tau=\mathcal{G}(r_h),
\end{equation}
then we define a characterized complex function $\mathcal{R}(z)$ related with the generalized free energy,
\begin{equation}\label{cf}
\mathcal{R}(z)\equiv\frac{1}{\tau-\mathcal{G}(z)},
\end{equation}
where we substitute the real variable $r_h$, {\it i.e.}, the event horizon radius, with a complex variable $z$. It means that we have extended to the complex plane $\mathbb{C}$ with a complex number defined by $z=x+ i y$ with $x, y\in \mathbb{R}$. We will see that the defect points previously defined by $\phi=0$ [cf. (\ref{phid})] on the real plane now turn into isolated singular points of the characterized complex function $\mathcal{R}(z)$ on the complex plane. 

\begin{figure}[htpb!]
\begin{center}
\includegraphics[width=5.3in,angle=0]{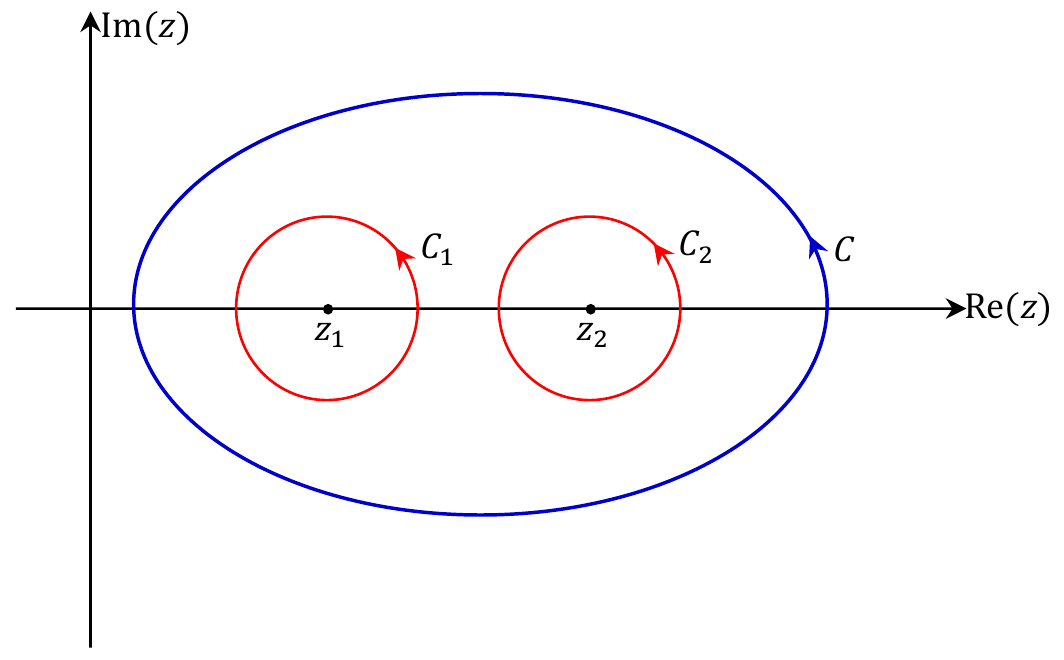}
\end{center}
\vspace{-5mm}
 \caption {A schematic diagram showing the contour integral of a function which is analytic except at finite isolated singular points $z_i$ along paths $C$ and $C_i$.}\label{fig1x}
\end{figure}

Then as shown in Fig. \ref{fig1x}, every real singular point, being a thermodynamic defect, can be endowed with an arbitrary contour $C_i$ enclosing it (not enclosing other singular points). The complex function $\mathcal{R}(z)$ is analytic on and within the contour $C_i$ except at the singular point. Then we can use the residue theorem to calculate the integral of the function $\mathcal{R}(z)$ along the path $C_i$ enclosing $z_i$. And according to the Cauchy-Goursat theorem [see (\ref{caug}) below], the integral of $\mathcal{R}(z)/2\pi i$ winding the singular points along the contour $C$ is equivalent to the integrals of $\mathcal{R}(z)$ along $C_i$ in the interior enclosing $z_i$.

This inspires us to reflect the winding number by using the residue of the complex function $\mathcal{R}(z)$ at its pole point, which is at the same time the topological defect corresponding to an on-shell black hole solution. To extract the information of the topology for the local defect point, we just need to normalize the residue, obtaining the winding number $w_i$ of a singular point $z_i$ as
\begin{equation}\label{wind}
w_i =\frac{\textrm{Res}\mathcal{R}(z_i)}{|\textrm{Res}\mathcal{R}(z_i)|}=\textrm{Sgn}[\textrm{Res}\mathcal{R}(z_i)],
\end{equation}
where $|\,|$ means to get the absolute value of the complex function, and $\textrm{Sgn}(x)$ stands for the sign function, which returns the sign of the real number and it gives $0$ if $x=0$. Here we should point out that as the singular points we care about are real, so we have $\textrm{Res}\mathcal{R}(z_i)\in\mathbb{R}$. Then the global topological number $W$ of the black hole spacetime can be obtained as
\begin{equation}\label{topw}
W=\sum_i w_i.
\end{equation}
This can be explained by applying the Cauchy-Goursat theorem, which enables us to calculate an integral of the complex function around an exterior contour through  integrals of simple interior closed contours enclosing singular points. So for the case shown in Fig. \ref{fig1x}, we have
\begin{equation}\label{caug}
\oint_C \mathcal{R}(z) d z= \sum_i \oint_{C_i} \mathcal{R}(z) d z.
\end{equation}

Calculating the winding number for the thermodynamic defects through residues enables us to deal with the different kinds of isolated singular points and discern them. This especially makes it possible to calculate the winding number of a pole of order above 1, which may be a generation point, annihilation point, or critical point. To see how these work, we will exemplify them in what follows by calculating the thermodynamic topologies of typical black hole solutions of Einstein equations (\ref{eineq}).


\section{Thermodynamic topologies of black holes}\label{sec:ttbh}

\subsection{Schwarzschild black hole}

We set the four-space metric as
\begin{equation}\label{met}
d s^2={-f(r) d t^2+ \frac{d r^2}{f(r)} } +r^2 d \theta^2+r^2 \sin ^2 \theta d \phi^2 .
\end{equation}
For a Schwarzschild black hole, $f(r)=1-\frac{2M}{r}$. The event horizon is at $r_h=2M$. The thermodynamic energy and entropy  of the system are respectively
\begin{equation}
E=M,
\end{equation}
\begin{equation}
S=4 \pi M^2.
\end{equation}
For the Schwarzschild black hole, according to the definition (\ref{cf}), we can explicitly obtain a characterized rational complex function 
\begin{equation}
\mathcal{R}_S(z)=\frac{1}{\tau-4\pi z},
\end{equation}
which is analytic away from an isolated pole of order 1 point
\begin{equation}
z_1=\frac{\tau}{4\pi}.
\end{equation}
According to the residue theorem, if we choose a contour $C_1$ around the pole $z_1$ [cf. Fig. \ref{fig1x}], then we will have
\begin{equation}
\textrm{Res}\mathcal{R}_S (z_1)=-\frac{1}{4\pi},
\end{equation}
which is non-zero. Otherwise, if we choose another contour, say $C_2$ which does not enclose $z_1$  in Fig. \ref{fig1x}, we will obtain
\begin{equation}
\oint_{C_2} \frac{\mathcal{R}_S(z)}{2\pi i} d z= 0.
\end{equation}
Obviously, according to Eqs. (\ref{wind}) and (\ref{topw}), we have 
\begin{equation}
w_0=-1, W=-1
\end{equation}
for the Schwarzschild black hole solution.


\subsection{Reissner-Nordström black hole}

For the Reissner-Nordström (RN) black hole, we have
\begin{equation}
f(r)=1-\frac{2M}{r}+\frac{Q^2}{r^2}
\end{equation}
in the metric (\ref{met}), where $Q$ is the electric charge. The generalized off-shell free energy of the system is
\begin{equation}
\mathcal{F}=\frac{ r_h^2+ Q^2}{2 r_h}-\frac{\pi  r_h^2}{\tau }.
\end{equation}
According to the definition (\ref{cf}), the characterized rational complex function can be written as
\begin{equation}\label{rrn}
\mathcal{R}_{RN}(z)=\frac{Q^2-z^2}{\tau  \left(Q^2-z^2\right)+4 \pi  z^3}.
\end{equation}
Applying the fundamental theorem of algebra,\footnote{Any polynomial
$$
p(x)=a_0+a_1 x+\cdots+a_n x^n, \quad a_n \neq 0
$$
can be factored completely as
$$
p(x)=a_n\left(x-z_1\right)\left(x-z_2\right) \cdots\left(x-z_n\right),
$$
where the $z_i$ are roots (which may be either complex or real).} 
we can write Eq. (\ref{rrn}) as
\begin{equation}
\mathcal{R}_{RN}(z)=\frac{Q^2-z^2}{4\pi (z-z_1)(z-z_2)(z-z_3)}.
\end{equation}
Root analysis shows that, for $0<\tau<6\sqrt{3}\pi Q$, we have only one real negative root, say $z_3$, so the on-shell condition cannot be satisfied. For the case of $\tau>6\sqrt{3}\pi Q$, we have \footnote{for $\tau\gg 1$, $z_1/Q\to 1^+$, and $z_1\ll z_2$. The extremal RN black hole corresponds to the case $\tau\to\infty$. Note that the condition $z_1>Q$ is important for  the calculation of the residue.}
\begin{equation}
0<z_1<z_2\in\mathbb{R},\, \textrm{and}\,\, 0>z_3\in\mathbb{R}.
\end{equation}
Then  according to the definition of the winding number Eq. (\ref{wind}), we have
\begin{equation}\label{rnr3}
w_1=1,\, w_2=-1.
\end{equation}
Thus according to Eq. (\ref{topw}), the topological number of the  RN black hole is 
\begin{equation}
W=0.
\end{equation}

What we should pay special attention to is the case $\tau=6\sqrt{3}\pi Q$, for which we have 
\begin{equation}
0<z_1=z_2\in\mathbb{R},\, \textrm{and}\,\, 0>z_3\in\mathbb{R}.
\end{equation}
So we get a second-order pole of the  function $\mathcal{R}_{RN}(z)$, for which we denote as $z_{12}$. Then the characterized function can be  explicitly written as
\begin{equation}
\mathcal{R}_{RN}(z)=\frac{Q^2-z^2}{4\pi (z-z_{12})^2 (z-z_3)}.
\end{equation}
$z_{12}$ is  a generation point as has been pointed out in Ref. \cite{Wei:2022dzw}. Here we can also give the winding number at this special point by calculating the residue of the analytic function at this pole of order 2 as
\begin{equation}\label{rnr2}
w_{1,2}=-1.
\end{equation}
This is a different result from  \cite{Wei:2022dzw}, where the winding number of this generation point is $0$. This means that the winding number of the generation point is definition-dependent. So does it for the annihilation point as we will see below.


\subsection{RN-AdS black hole}
For the RN black hole with a negative cosmological constant $\Lambda$, we have the metric function in Eq. (\ref{met}) as
\begin{equation}\label{metads}
f(r)=1-\frac{2M}{r}+\frac{Q^2}{r^2}+\frac{r^2}{L^2},
\end{equation}
where $L$ is the AdS length scale, relating with the negative cosmological constant through a relation $\Lambda=-3/L^2$. The generalized off-shell free energy of the system is
\begin{equation}\label{rrnx}
\mathcal{F}=\frac{-\Lambda r_h^4+3 r_h^2+3 Q^2}{6 r_h}-\frac{\pi r_h^2}{\tau }.
\end{equation}
 According to the definition (\ref{cf}), we have the generalized characterized complex function
\begin{equation}\label{rrn2}
\mathcal{R}_{RN-AdS}(z)=\frac{Q^2+\Lambda z^4-z^2}{\tau \left(Q^2+\Lambda z^4-z^2\right)+4 \pi z^3}.
\end{equation}
 Based on the fundamental theorem of algebra as aforementioned, Eq. (\ref{rrn2}) can be rewritten as
\begin{equation}
\mathcal{R}_{RN-AdS}(z)=\frac{Q^2+\Lambda z^4-z^2}{\tau\Lambda (z-z_1)(z-z_2)(z-z_3)(z-z_4)}\equiv\frac{Q^2+\Lambda z^4-z^2}{\mathcal{A}(z)},
\end{equation}
where we have defined a polynomial function $\mathcal{A}(z)$ for brevity.

\begin{figure}[htpb!]
\begin{center}
\includegraphics[width=5.3in,angle=0]{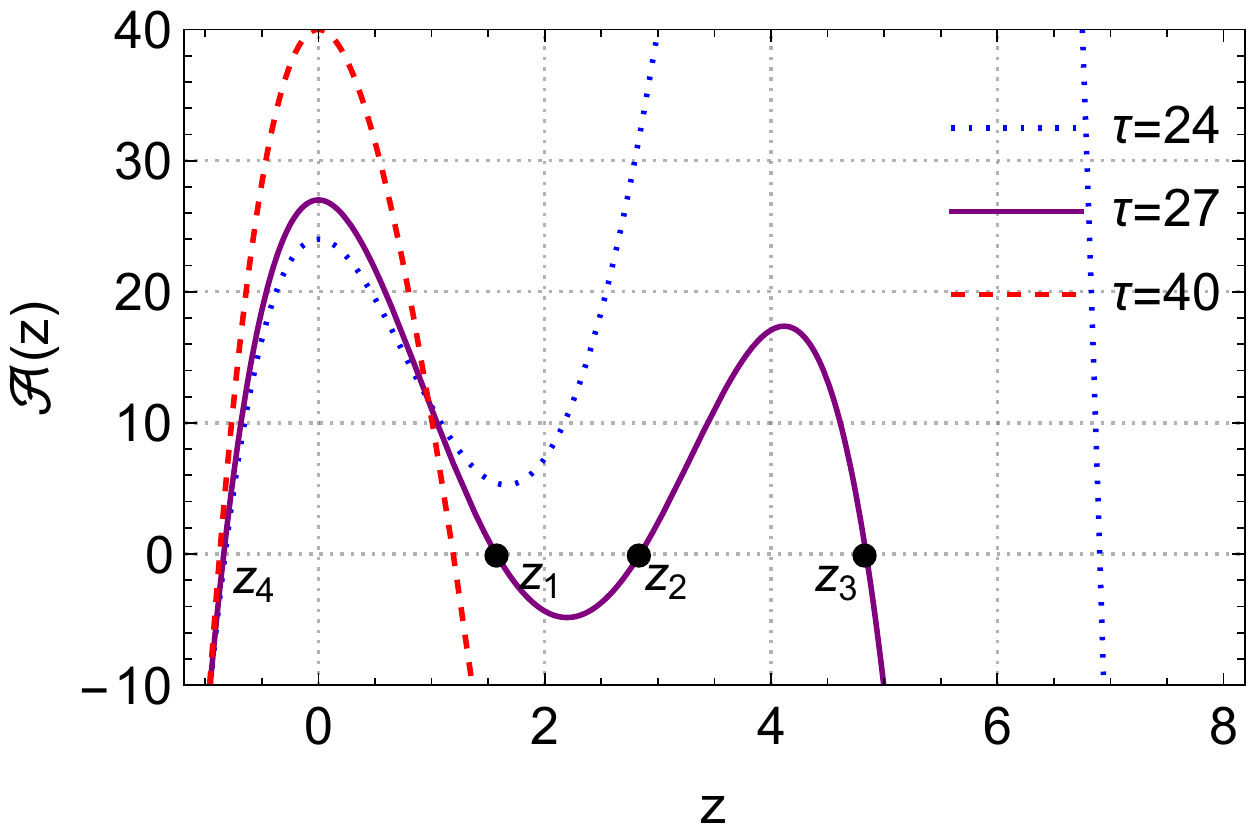}
\end{center}
\vspace{-5mm}
 \caption {Roots of the polynomial equation $\mathcal{A}(z)=0$ with $Q=1,\,\Lambda=-8\pi\times 0.0022$.}
 \label{root_rnads}
\end{figure}

First, we focus on the case where there are four distinct real roots. One of the roots, suppose it is $z_4$, is negative, and another three real roots satisfy the relation $z_1<z_2<z_3$, as shown in Fig. \ref{root_rnads}. As the fourth term in Eq. (\ref{metads}) is positive, we can deduce that $r_h>M>Q$, or else there will be no horizon. So we have $z_1>Q$. In this case, we have the local winding numbers 
\begin{equation}
w_1=1,\,w_2=-1,\,w_3=1
\end{equation}
by using Eq. (\ref{wind}). Then we obtain 
\begin{equation}
W=w_1+w_2+w_3=1
\end{equation}
for the global topological number of the RN-AdS black hole.

Notice that in the above analysis we do not need to give a specific range of the value of $\tau$ and the residue can be calculated under the condition $\tau>0$. However, in fact there should be a range for $\tau\in (\tau_{\textrm{b}},\,\tau_{\textrm{t}})$ where there are four real roots. 

For $\tau<\tau_{\textrm{b}}$, as shown in Fig. \ref{root_rnads}, there is only one real positive root, say, $z_3$, then we have the local winding number of that point, 
\begin{equation}
w_3=1,
\end{equation}
whist the global topological number of the black hole is 
\begin{equation}
W=1.
\end{equation}

For $\tau>\tau_{\textrm{b}}$, as shown in Fig. \ref{root_rnads}, there is still only one real positive root, say, $z_1$. Then we have the local winding number of that point being
\begin{equation}
w_1=1,
\end{equation}
and the global topological number is still
\begin{equation}
W=1.
\end{equation}

So we can know that the topological number of the RN-AdS black hole is always $W=1$.

For $\tau=\tau_{\textrm{b}}, \tau_{\textrm{t}}$, we will have $z_1=z_2$, and $z_2=z_3$, respectively. We denote the two merging points by $z_{1,2}$ and $z_{2,3}$, which are individually the generation point and annihilation point. For the former case, we have 
\begin{equation}
\mathcal{R}_{RN-AdS}(z)=\frac{Q^2+\Lambda z^4-z^2}{\tau\Lambda (z-z_{1,2})^2 (z-z_3)(z-z_4)},
\end{equation}
where $z_{1,2}$ is a root of multiplicity 2.
Then by Eq. (\ref{wind}), we should calculate the residue of a pole of order 2, yielding the winding number at this point as
\begin{equation}
w_{1,2}=-1.
\end{equation}
For the latter case, we have 
\begin{equation}
\mathcal{R}_{RN-AdS}(z)=\frac{Q^2+\Lambda z^4-z^2}{\tau\Lambda (z-z_{1}) (z-z_{2,3})^2(z-z_4)},
\end{equation}
by which we should calculate the second-order pole point $z_{2,3}$, which yields
\begin{equation}
w_{2,3}=1.
\end{equation}

Combining with the result for RN black hole in Eq. (\ref{rnr2}), we see that the residue method gives different local topological properties for the generation point and annihilation point, each with a negative and positive winding number, respectively. The topological current method used in \cite{Wei:2022dzw} gives winding number 0 for both the two kinds of points, since the necessity of current conservation.

There is another possibility of $z_1=z_2=z_3$, which corresponds to the critical case and we just name the root of multiplicity 3 as a critical point. For consistency of analysis together with the case of black holes in non-linear electrodynamics below, we will discuss it in the closing remarks [cf. Eq. (\ref{z1232})].


\subsection{Black holes in non-linear electrodynamics}
Now we consider the black hole solutions in non-linear electrodynamics with an action 
\begin{equation}
S=\int d^4 x \sqrt{-g}\left(R-2 \Lambda-\sum_{i=1}^{\infty} \alpha_i\left(F^2\right)^i\right),
\end{equation}
where $\alpha_i$ are dimensional coupling constants and $F^2\equiv F_{\mu\nu}F^{\mu\nu}$ with $F_{\mu\nu}$ the electromagnetic tensor. The black hole solution is \cite{Gao:2021kvr,Tavakoli:2022kmo}
\begin{equation}
\begin{aligned}
d s^2 &=-U(r) d t^2+\frac{1}{U(r)} d r^2+r^2 d \Omega_2^2, \\
A_\mu &=[\Phi(r), 0,0,0],
\end{aligned}
\end{equation}
where
\begin{equation}
\Phi=\sum_{i=1}^{\infty} b_i r^{-i}, \quad U=1+\sum_{i=1}^{\infty} c_i r^{-i}+\frac{r^2}{l^2}.
\end{equation}
For $\alpha_i \neq 0$ with $i \leq 7$, we can write them explicitly as
\begin{equation}
\Phi=\frac{Q}{r}+\frac{b_5}{r^5}+\frac{b_9}{r^9}+\frac{b_{13}}{r^{13}}+\frac{b_{17}}{r^{17}}+\frac{b_{21}}{r^{21}}+\frac{b_{25}}{r^{25}},
\end{equation}
\begin{equation}
\begin{aligned}
U=&-\frac{2 M}{r}+\frac{Q^2}{r^2}+\frac{b_5 Q}{2 r^6}+\frac{b_9 Q}{3 r^{10}}+\frac{b_{13} Q}{4 r^{14}} \\
&+\frac{b_{17} Q}{5 r^{18}}+\frac{b_{21} Q}{6 r^{22}}+\frac{b_{25} Q}{7 r^{26}}+\frac{r^2}{l^2}
\end{aligned}
\end{equation}
with $b_i=0$ for $i>25$. With $\alpha_1=1$, we have
\begin{equation}
c_1=-2 M \quad c_i=\frac{4 Q}{i+2} b_{i-1}, \quad \text { for } \quad i>1,
\end{equation}
where
\begin{equation}
b_1=Q, \quad b_5=\frac{4}{5} Q^3 \alpha_2, \quad b_9=\frac{4}{3} Q^5\left(4 \alpha_2^2-\alpha_3\right),
\end{equation}
\begin{equation}
b_{13}=\frac{32}{13} Q^7\left(24 \alpha_2^3-12 \alpha_3 \alpha_2+\alpha_4\right),
\end{equation}
\begin{equation}
b_{17}=\frac{80}{17} Q^9\left(176 \alpha_2^4-132 \alpha_2^2 \alpha_3+16 \alpha_4 \alpha_2+9 \alpha_3^2-\alpha_5\right),
\end{equation}
\begin{equation}
\begin{aligned}
b_{21}=& \frac{64}{7} Q^{11}\left(1456 \alpha_2^5+234 \alpha_3^2 \alpha_2+208 \alpha_4 \alpha_2^2-24 \alpha_4 \alpha_3\right.\\
&\left.-1456 \alpha_2^3 \alpha_3-20 \alpha_5 \alpha_2-\alpha_6\right),
\end{aligned}
\end{equation}
\begin{equation}
\begin{aligned}
b_{25}=&\frac{448}{25} Q^{13}\left(13056 \alpha_2^6+2560 \alpha_2^3 \alpha_4-720 \alpha_3 \alpha_2 \alpha_4\right. \\
&+16 \alpha_4^2-300 \alpha_2^2 \alpha_5+4320 \alpha_3^2 \alpha_2^2-16320 \alpha_2^4 \alpha_3 \\
&\left.-24 \alpha_6 \alpha_2-135 \alpha_3^3+30 \alpha_3 \alpha_5-\alpha_7\right).
\end{aligned}
\end{equation}

\begin{figure}[htpb!]
\begin{center}
\includegraphics[width=5.5in,angle=0]{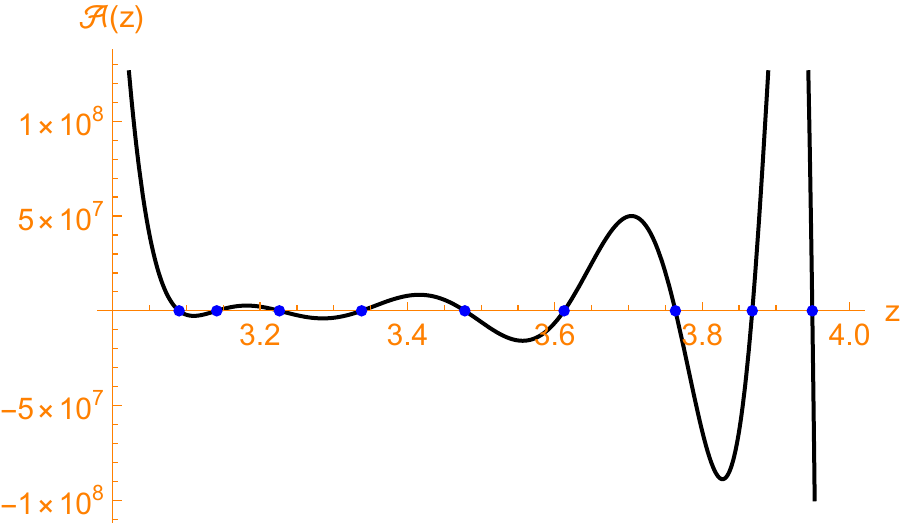}
\end{center}
\vspace{-5mm}
 \caption {Positive real roots of the polynomial equation $\mathcal{A}(z)=0$ with $P=0.0006737,\,Q=2.4544565,\,\alpha_2=-5.3439180,\,\alpha_3=72.4523175,\,\alpha_4=-1248.4241037,\,\alpha_5=23241.5416893,\,\alpha_6=431538.2300431,\,\alpha_7=7.4168150\times 10^6,\,\tau=52.2102945$. We denote the roots as $z_1=3.0901936,\,z_2=3.1412747,\,z_3=3.2262569,\,z_4=3.3377906,\,z_5=3.47789817,\,z_6=3.6125760,\,z_7=3.7636139,\,z_8=3.8678889,\,z_9=3.9493271 $ which correspond to blue points from left to right in the diagram. Besides, there is one another real root not shown whose locus is $z_{10}=-2.3623414$.}
 \label{root_ne}
\end{figure}

 The entropy and mass of the black hole is \cite{Tavakoli:2022kmo}
\begin{equation}
S=\pi r_{h}^2,
\end{equation}
\begin{equation}
M=\frac{b_5 Q}{4 r_h^5}+\frac{b_9 Q}{6 r_h^9}+\frac{b_{13} Q}{8 r_h^{13}}+\frac{b_{17} Q}{10 r_h^{17}}+\frac{b_{21} Q}{12 r_h^{21}}+\frac{b_{25} Q}{14 r_h^{25}}+\frac{4}{3} \pi P r_h^3+\frac{Q^2}{2 r_h}+\frac{r_h}{2}.
\end{equation}
Then we have the generalized off-shell free energy of the system as
\begin{equation}
\mathcal{F}=M-\frac{S}{\tau},
\end{equation}
which yields a characterized rational function \footnote{One subtlety here is that if we do a transformation $$\mathcal{A}(z)\to -\mathcal{A}(z),$$ then both the winding number and topological number will change sign. We here choose the sign of $\mathcal{A}(z)$ as it can be reduced to the RN-AdS case in the vanishing $\alpha_{i>1}$ limit.}
\begin{equation}
\mathcal{R}_{NE}(z)\equiv\frac{\mathcal{B}(z)}{\mathcal{A}(z)},
\end{equation}
where
\begin{equation}
\begin{aligned}
\mathcal{A}(z)=&-350 b_5 Q \tau z^{20}-420 b_9 Q \tau z^{16}+1120 \pi P \tau z^{28}-140 Q^2 \tau z^{24}-560 \pi z^{27}+140 \tau z^{26}\\&-500 b_{25} Q \tau -455 b_{13} Q \tau z^{12}-476 b_{17} Q \tau z^8-490 b_{21} Q \tau z^4,
\end{aligned}
\end{equation}
\begin{equation}
\begin{aligned}
\mathcal{B}(z)=&-350 b_5 Q z^{20}-420 b_9 Q z^{16}-455 b_{13} Q z^{12}-476 b_{17} Q z^8-490 b_{21} Q z^4-500 b_{25} Q\\&+1120 \pi P z^{28}-140 Q^2 z^{24}+140 z^{26}.
\end{aligned}
\end{equation}
As explained in the captain of Fig. \ref{root_ne}, we can calculate the local winding number of each positive real root by Eq. (\ref{wind}), 
\begin{equation}
w_{2k-1}=1,\,w_{2k}=-1,\,w_9=1,\quad k=1, 2, 3, 4.
\end{equation}
As a result, we have the global topological number
\begin{equation}
W=\sum_{i=1}^{9}w_i=1.
\end{equation}
This is the same with the RN-AdS black hole case. However, there will be a different number of generation/annihilation points. If we denote $w_{i,j}$ as the merging of roots $z_i$ and $z_j$, we will have merging points $w_{i,i+1}, (i=1,2,\ldots,8) $, and the winding number of them are 
\begin{equation}
w_{2k-1,2k}=-1, w_{2k,2k+1}=1, k=1,2,3,4.
\end{equation}
We can identify them individually as generation points and annihilation points.


\section{Closing remarks}\label{sec:con}
In this paper, we proposed a new formulation to study the thermodynamic topology properties of black holes. In the generalized free energy landscape, we defined a characterized complex function (\ref{cf}) based on the off-shell free energy of the black hole. Instead of applying Duan's topological current theory \cite{Wei:2022dzw}, we perform contour integrals around singular points of the defined complex function, which according to the residue theorem can be conducted by calculating the residues of those singular points. Local topological properties are closely related to the winding numbers of isolated poles of order 1, which can be conducted by applying (\ref{wind}). With our definition, positive and negative winding numbers stand for a locally stable black hole with positive specific heat and a locally unstable black hole with negative specific heat. The global topological number of a black hole, which can be used to discern spacetime categories,   can be obtained by adding all winding numbers of each singular point. We have exemplified our residue method by studying winding numbers and topological numbers for Schwarzschild black holes, RN black holes, and RN-AdS black holes. Obtained results are consistent with those in \cite{Wei:2022dzw}. Moreover, we have also studied the thermodynamic topological properties of black holes in non-linear electrodynamics and found that winding numbers of locally stable (unstable) black hole branches are always positive (negative). And we also found that the non-linear electromagnetic field does not change the global topological properties of the AdS black hole, holding its topological number the same as that of the RN-AdS black hole.

Surprisingly, the representation of topological properties for the generation and annihilation points can be quite different upon using Duan's topological current method and the residue method. The former one yields a winding number 0 for both generation and annihilation points whilst the latter one gives a winding number -1 for the generation point and 1 for the annihilation point, respectively. Duan's topological current is conserved so a 0 winding number at both the generation and annihilation points is inevitable. However, in the residue method, the generation and annihilation points are roots of multiplicity 2. So when calculating the residue of such poles of order 2, the derivatives of the characteristic function constructed from the off-shell free energy are involved. Due to the sign of the derivative of the characteristic function of a generation point being contrary to that of an annihilation point, it is not difficult to understand that we get different winding numbers for the two kinds of points. It is this different representation of the topological properties for the generation and annihilation points that makes the residue method an effective complement of Duan's topological current method.

One may notice that for the RN-AdS black hole, there may be a merging of three roots $z_1,\,z_2,\,z_3$. Indeed, for $\tau=3 \sqrt{6} \pi Q,\, \Lambda=-1/12 Q^2$, we do have this merging point $z_{1,2,3}=\sqrt{6}Q$. We then obtain 
\begin{equation}\label{z123}
\textrm{Res}\mathcal{R}_{RN-AdS} (z_{1,2,3})=\frac{7}{8\pi},
\end{equation}
independent of the black hole charge $Q$. In this case, according to the definition of the winding number (\ref{wind}), we have
\begin{equation}\label{z1232}
w_{1,2,3}=\textrm{Sgn}[\textrm{Res}\mathcal{R}(z_{1,2,3})]=1.
\end{equation}
This result is also different from the one in \cite{Wei:2022dzw}, where the winding number is 0 for a critical point (also due to the conservation of the topological current). However, one question arises: if we consider the non-linear electromagnetic black hole, which admits multi-critical points \cite{Tavakoli:2022kmo,Wu:2022bdk} and then more than three points in Fig. \ref{root_ne} will merge, then one may ask what the local topological properties are for those different multi-merging points. We believe this question deserves further investigation. 

There is one other question: might we distinguish standard defect points, generation/annihilation points, and critical points? This question arises as the winding numbers of these points only take values $-1$ and $1$; and the answer is yes. To illustrate how this can be conducted, let us expand the characterized complex function $\mathcal{R}(z)$ defined in (\ref{cf}) at its singular point $z_0$ as a Laurent series
\begin{equation}
\mathcal{R}(z)=\sum_{n=-\infty}^{\infty} a_n\left(z-z_0\right)^n,
\end{equation}
where 
\begin{equation}
a_n=\frac{1}{2 \pi i} \oint_C \frac{f(\xi)}{\left(\xi-z_0\right)^{n+1}} d \xi
\end{equation}
with $C$ a contour encircling $z_0$. It is well-known that $\operatorname{Res} \mathcal{R}(z_0)=a_{-1}.$ Then immediately we can find that: for a standard defect point, 
\begin{equation}
a_{n<-1}=0, a_{n\geq-1}\neq 0;
\end{equation}
for a generation or annihilation point, we have  
\begin{equation}
a_{n<-2}=0, a_{n\geq -2}\neq 0;
\end{equation}
for a critical point, we have 
\begin{equation}
a_{n<-3}=0, a_{n\geq -3}\neq 0.
\end{equation}
Consequently, all those points can be discerned.

Lastly, note that we mainly focused on  the application of residue method on studying the thermodynamic topologies of stationary  spherically symmetric black holes. However, the method can be easily generalized to stationary cases \cite{Wu:2022whe}.

\section*{Acknowledgements}
We are grateful to Shao-Wen Wei for useful discussions. We also acknowledge the anonymous referee for a constructive suggestion. M. Z. is supported by the National Natural Science Foundation of China with Grant No. 12005080.  J. J.  is supported by the National Natural Science Foundation of China with Grant No. 12205014, the Guangdong Basic and Applied Research Foundation with Grant No. 217200003, and the Talents Introduction Foundation of Beijing Normal University with Grant No. 310432102.

\bibliographystyle{jhep}
\bibliography{refs}
\end{document}